\documentclass{agujournal}

\usepackage{booktabs}
\usepackage{hyperref}

\begin{document}

\title{Earthquake catalog-based machine learning identification of laboratory fault states and the effects of magnitude of completeness.}

\authors{Nicholas Lubbers\affil{1}, David C. Bolton\affil{2}, Jamaludin Mohd-Yusof\affil{3}, Chris Marone\affil{2}, Kipton Barros\affil{1}, Paul A. Johnson\affil{4}}

\affiliation{1}{Theoretical Division and CNLS, Los Alamos National Laboratory, Los Alamos, New Mexico, USA}
\affiliation{2}{Department of Geosciences, Pennsylvania State University, University Park, Pennsylvania, USA}
\affiliation{3}{Computer, Computational, and Statistical Sciences Division, Los Alamos National Laboratory, Los Alamos, New Mexico, USA}
\affiliation{4}{Geophysics Group, Los Alamos National Laboratory, Los Alamos, New Mexico, USA}

\correspondingauthor{Paul A. Johnson}{paj@lanl.gov}

\begin{abstract}

Machine learning regression can predict macroscopic fault properties such as shear stress, friction, and time to failure using continuous records of fault zone acoustic emissions. Here we show that a similar approach is successful using event catalogs derived from the continuous data. Our methods are applicable to catalogs of arbitrary scale and magnitude of completeness. We investigate how machine learning regression from an event catalog of laboratory earthquakes performs as a function of the catalog magnitude of completeness. We find that strong model performance requires a sufficiently low magnitude of completeness, and below this magnitude of completeness, model performance saturates.

\end{abstract}

\section{Introduction}

Earthquake catalogs are a key seismological tool for problems involving location, precursors, and earthquake source parameters \citep{Gutenberg56,Scholz68,AkiRichards,Meier2017}. Catalogs are fundamental to determining where, when, and how a fault has ruptured \citep{AkiRichards,Scholz2002} and they are central to studies of earthquake precursors and seismic hazard based on both field observations \citep{Marsan2014,Bouchon2008,Bouchon2013,Huang2017,Meng2018} and theory \citep{AkiRichards,Ferdowsi2013,Kazemian}. Laboratory studies of earthquake precursors also rely on event catalogs \citep{Lei2014,JohnsonPrecursor,riviere2018}. Such studies include the temporal evolution of the Gutenberg-Richter b-value, which is a fundamental parameter relating laboratory and field studies of earthquake physics \citep{Smith1980,Wang2016,Scholz2015,Gulia,riviere2018,Kwiatek2014}. The seismic b-value has emerged as a key parameter for estimating seismic hazard and for connecting laboratory and field studies of seismicity \citep{Scholz2002}.

Laboratory-based studies of seismic hazard and earthquake forecasting have traditionally relied on earthquake catalogs. However, recent work shows that lab earthquakes can be predicted based on continuous acoustic data \citep{rouet17ML,BRLFriction,Hulbert2017SlowSlip}. These works show that statistical characteristics of the continuous seismic signal emanating from lab fault zones can predict the timing of future failure events as well as the frictional state of the fault zone. Moreover, recent work has extended this approach to field observation by showing that statistical characteristics of continuous seismicity recorded in Cascadia contain a fingerprint of the megathrust displacement rate that can be used to predict the timing of episodic tremor and slip events \citep{RouetCascadia18}. The methods are based on machine learning (ML) and rely exclusively on continuous measurements of acoustic emission (AE). Thus, a key question involves whether this approach can be extended to catalog based measurements of seismic activity.

Applications of ML to seismology and geosciences are becoming increasingly common \citep{lary2016,khawaja18,kenta18,zefeng18,Holtzman18}. Here, we apply the ML Random Forest method \citep{Breiman2001} to study laboratory earthquake catalogs.  We find that as the catalog degrades, so does our ability to infer fault physics, as quantified the ability to predict shear stress and the time to and since failure.  We begin by describing the biaxial shearing experiment, our methods to construct a catalog from the continuous waveform data, the statistical descriptors we extract from the catalog, and the random forest method used to make inferences from these descriptors. This results in successful models for estimating the fault physics variables using local-in-time catalog data.  Then, from the original catalog we construct a series of progressively more degraded catalogs, each of which contains only the events that exceed a variable magnitude of completeness.  We examine the performance of this method as a function of the magnitude of completeness, and observe that the learned signatures of fault physics degrade dramatically if small enough events are not detected. However, the performance plateaus for low magnitudes of completeness; the smallest 80\% events in the catalog can be dropped without sacrificing performance. Our results show that ML methods based on event catalogs can successfully model laboratory faults, with the caveat that predictive power requires a sufficiently complete catalog.

\section{Data Collection}

\subsection{Biaxial Shear Experiment}

\begin{figure}[h]
 \centering
 \includegraphics[width=32pc]{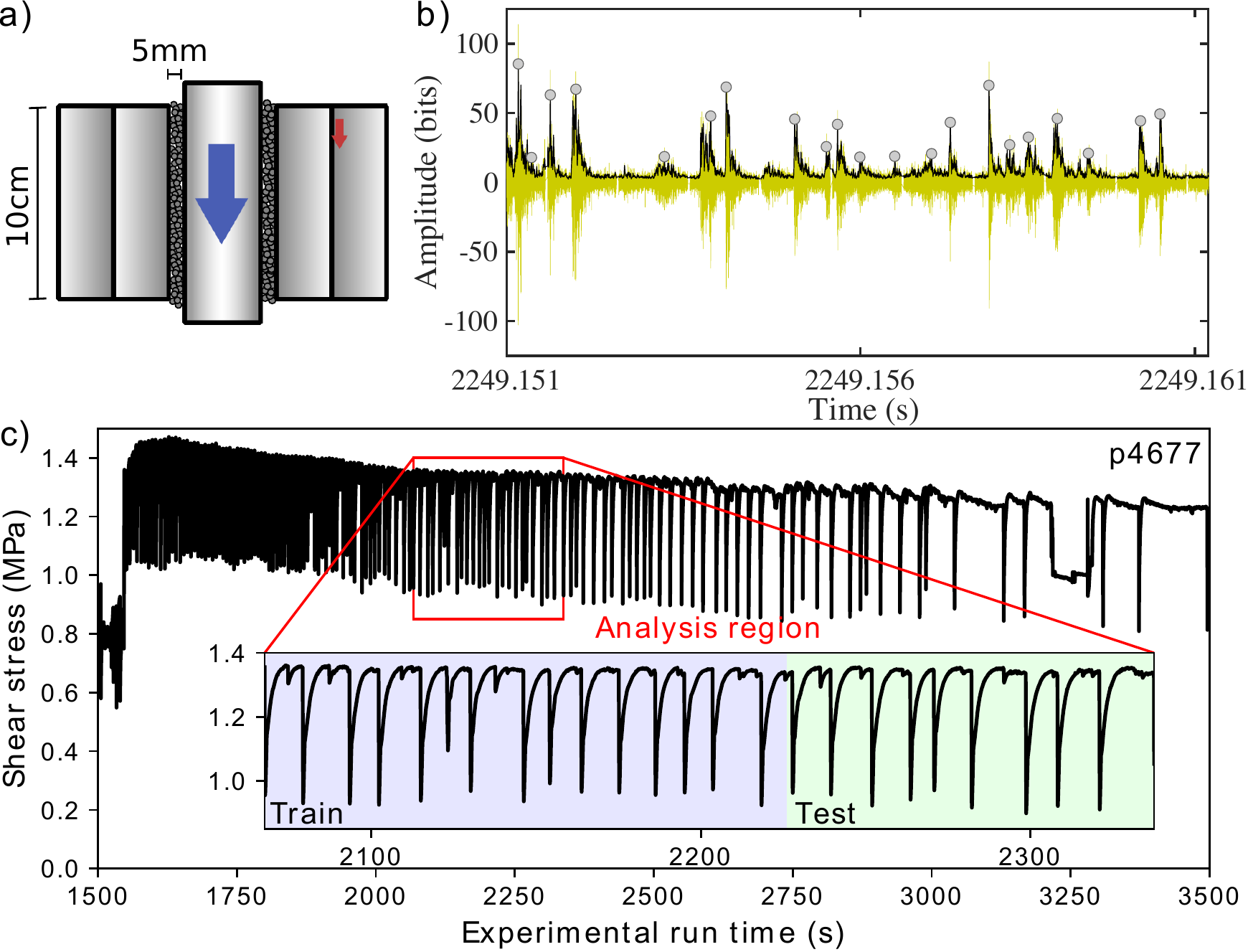}

 \caption{{\bf a)} Diagram of biaxial shear apparatus. The blue arrow indicates the direction of shear, and the red arrow locates the piezoceramic sensor that records AE. {\bf b)} Example cataloged events (grey circles) derived from continuously recorded AE (yellow) using the statistics of the smoothed signal envelope (black). {\bf c)} Observed shear stress for an experiment at fixed strain rate. Sharp drops in stress correspond to failure events. Inset: The data segment analyzed in this paper. An event catalog is constructed from AE data sampled at 4 MHz. Our models are constructed using event data from the training segment (blue) and evaluated on data from the test segment (green). 
 }
 \label{figstress}
 \end{figure}

We used data from laboratory friction experiments conducted with a biaxial shear apparatus \citep{marone98,JohnsonPrecursor} pictured in Fig.~\ref{figstress}a. Experiments were conducted in the double direct shear configuration in which two fault zones are sheared between three rigid forcing blocks. Our samples consisted of two 5 mm-thick layers of simulated fault gouge with a nominal contact area of 10 x 10 cm$^2$. Gouge material consisted of soda-lime glass beads with initial particle size between 105-149 {\textmu}m. Additional details about the apparatus and sample construction can be found in \citet{Anthony05}, \citet{riviere2018}, and Text S1.  Prior to shearing, we impose a constant fault normal stress of 2 MPa using a servo-controlled load-feedback mechanism and allow the sample to compact. Once the sample has reached a constant layer thickness, the central block is driven down at constant rate of 10~\textmu m/s. In tandem, we collect an AE signal continuously at 4~MHz (Fig.~\ref{figstress}b, yellow curve) from a piezoceramic sensor embedded in a steel forcing block $\approx 22$ mm from the gouge layer. 

Figure~\ref{figstress}c shows the shear stress measured on the fault interface for a full experiment.  Experiments begin with a run-in stage where the shear stress increases and macroscopic shearing of the fault zone transitions from stable sliding to stick-slip failure. The repetitive cycles of loading and failure represent laboratory seismic cycles and transition from periodic to aperiodic as a function of load-point displacement in our experiments \citet{Anthony05,JohnsonPrecursor}. Here, we focus on aperiodic slip cycles and measure the slip history for a set of 23 failure events that occur over 268~seconds~\ref{figstress}c. We define large failure events as times for which stress drop exceeds 0.05 MPa within 1 ms.

\subsection{Event Catalog}
\label{sec:events}

\begin{figure}[h]
 \centering
 \includegraphics[width=32pc]{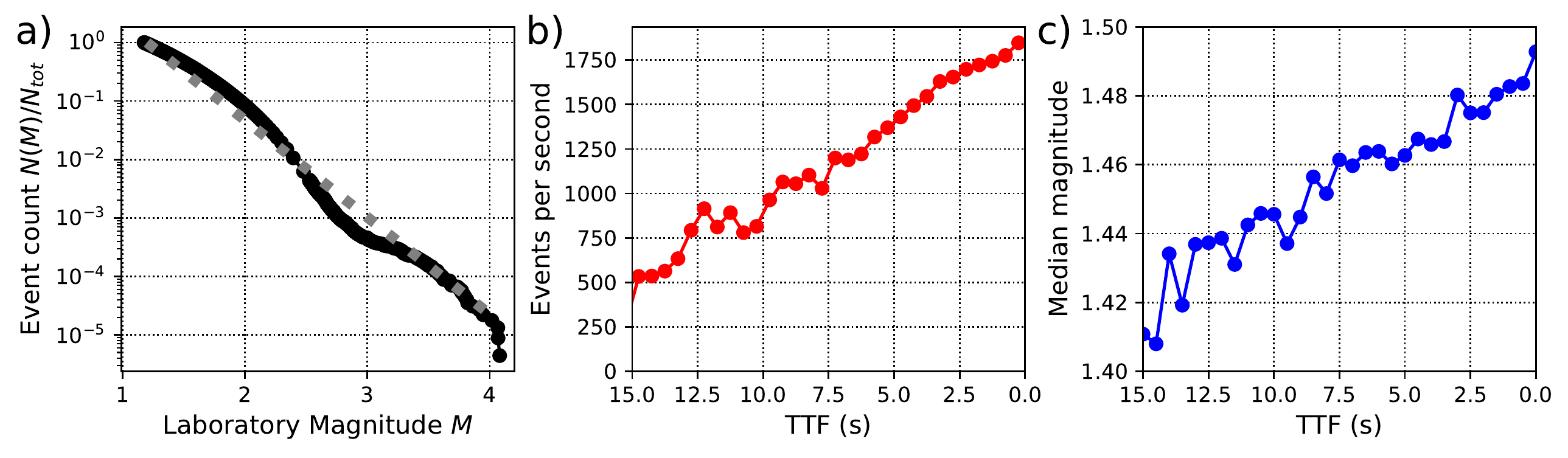}
 \caption{ {\bf a)} Normalized count of training events with magnitude exceeding $M$. The dashed line represents ideal Gutenberg-Richter scaling with a $b$-value of 1.66. {\bf b)} Event rates vs. TTF stacked over all 23 failure cycles. {\bf c)} Stacked event statistics for median event magnitude vs. time to failure.}
 \label{figstack}
 \end{figure}

Our event catalog is built from the acoustic signal (example in Fig.~\ref{figstress}b, yellow) during the time interval noted in Fig.~\ref{figstress}c, inset. The catalog is composed of event amplitudes  $A(t_k)$ and times $t_k$, where $k$ is an event index, built following methods described in \citet{riviere2018} and Text S2. We define laboratory magnitudes $m_k = \log_{10}{A(t_k)}$. The catalog contains  $N_\textrm{tot} \approx 3.8 \times 10^5$ slip events with laboratory magnitudes that range from about 1.0 to 4.2. Our lab magnitudes are based only on acoustic signal amplitude, and although they do not connect directly to earthquake magnitude, previous works have demonstrated the connection between seismic b-value and lab frequency magnitude statistics \citep{Scholz2002}.

To visualize the distribution of magnitudes, we construct the cumulative event distribution
\begin{equation}
N(M) = \sum_{m_k \geq M} 1. \label{eq:n_M}
\end{equation}That is, $N(M)$ is the count of all events with magnitude at least $M$. We plot $\log N(M) / N_\textrm{tot}$ in Fig.~\ref{figstack}a and observe reasonable Gutenberg-Richter scaling with a $b$-value of 1.66. 

Figure~\ref{figstack}b shows the rate of events as a function of time to failure (TTF), averaged over all slip cycles.  The rate of events increases significantly as failure is approached, similar to what has been observed in prior works \citep{JohnsonPrecursor,weeks1978,Ponomarev1997,riviere2018,Goebel2013}.  Figure~\ref{figstack}c shows that the median event magnitude changes steadily but slowly from approximately $M_{\textrm{median}}=1.4$ to $M_{\textrm{median}}=1.5$  as a function of time during the lab seismic cycle. The trends of the catalog event count and size reflected in Fig.~\ref{figstack}b and Fig.~\ref{figstack}c motivate the creation of machine learning features that describe the evolution event count and size in more detail.

\section{Machine Learning Methods}

\subsection{Regression with Random Forests}

Abstractly, a regression model maps input points $\mathbf{X}_i$ to continuous output labels  $\hat y_i$. Input points are represented by a feature vectors $\mathbf X_i = (X_{i,1}, X_{i,2},\dots X_{i, N_\textrm{features}})$ that are designed to capture the relevant information about $\hat y_i$. The training dataset $\{(\mathbf X_i, y_i)\}$ consists of many points $\mathbf X_i$ associated with true labels $y_i$. From this training set, one aims to learn a model that is successful at predicting labels $\hat y_i$ for {\em new} data points that were unseen within the training process. In this work, the features will be statistics extracted from the catalog, the labels will describe the macroscopic state of the fault (such as shear stress); both are indexed by time window $i$. 

We  use the random forest (RF) algorithm \citep{Breiman2001} as it is implemented in the scikit-learn package \citep{pedregosa2011scikit}. A single RF model contains many decision trees, each of which is a simple regression model constructed stochastically from the training data. To make a prediction, the decision tree begins at its root node and asks a series of yes/no questions. Each question has the form: {\em Is some input feature above or below some threshold value?} Eventually a unique leaf node is reached, from which the decision tree makes its prediction. The RF makes a prediction by averaging over many decision trees, and tends to be more robust than any individual tree. Additional details about our training procedure are available in Text S3. We investigated other regression methods \citep{Friedman2001,hui2005,tibshirani96,hoerl1970} besides the RF, and although all were successful to some extent, none offered definitive improvement over the RF (see Table~S1). 

A regression model is successful if its predicted labels $\hat y_i$ agree with the true labels $y_i$ for data points in the {\em testing dataset}, which is separate from the training dataset. We use the dimensionless $R^2$ value to quantify the performance of the model,

\begin{equation}
R^2 = 1 - \frac{\sum_i (\hat{y_i}-y_i)^2}{\sum_i (\bar{y}-y_i)^2},
\end{equation}
where $\bar{y}$ is the mean output label for the testing dataset. Let us review the significance of the $R^2$ metric, which is maximized at $R^2=1$, and can take on arbitrarily low values. Using the mean label as a constant predictor for all data points, $\hat y_i = \bar y$, yields $R^2=0$. Thus, $R^2<0$ indicates that a model performs worse than a constant learned only from the labels, without utilizing the features; if $R^2 \leq  0$, a model has failed to discover useful correlations between the features and the labels. The condition $R^2=1$ would indicate that a model makes perfect predictions for all labels in the test set.

\subsection{Feature and Label Creation}

Our task is to build a regression model that uses the event distribution within a {\em local} time window to understand and make predictions about the stress state and times to and from large failure events. Each time window is labeled, and predictions of the labels will depend only on features from within the window. Thus, each test prediction is independent of prior predictions, and the model cannot exploit the quasi-periodicity of the fault state.

There is significant interest in predicting the TTF for an upcoming earthquake event. \citet{rouet17ML} and \citet{BRLFriction} showed that the continuous AE from a fault is remarkably predictive of the TTF and instantaneous friction. In both previous works, $\mathbf X_i$ is a collection of statistical features extracted from the {\em continuous} AE signal for the time window. Here, we build features for each time window using only the event catalog, rather than the continuous acoustic signal. 

We divide our full event catalog (cf. Sec.~\ref{sec:events}) into sub-catalogs that contain only the events occurring in non-overlapping time windows of length $\Delta T = 1$ second. The feature vector contains information about events in the sub-catalog for this time window. Our analysis region consists of 270 seconds, which we divide into 1 second windows. We train on the first $60\%$ (159 windows) and test on the remaining $40\%$ (111 windows).

We construct regression models for three labels: (a) the shear stress, (b) the time to the next failure event (TTF), and (c) the time since the last failure event (TSF). The shear stress label is assigned using the mean shear stress over that window. TTF is measured following the end of time window, and TSF is measured preceding the beginning of the time window. For windows including failures, both TTF and TSF are zero.

Abstractly, our the features for each time window describe the cumulative statistics of event counts and amplitudes from the sub-catalog evaluated at magnitude thresholds extracted from the full catalog. More precisely, the feature vector $\mathbf X_i$ for time window $i$ will have components $X_{i,1}, X_{i,2}, \dots X_{i,N_\textrm{features}}$. Each feature component $X_{i,j}$  is associated with a characteristic magnitude $M_j$. For each $j$, we define $M_j$ to be the largest observed magnitude that satisfies
\begin{equation}
\frac{N(M_j)}{N_\textrm{tot}} \geq \alpha^j,
\end{equation}
where the $j$ in right hand side is applied as an exponent; $\alpha$ is a parameter of range $ 0 < \alpha <1$ that controls the fineness of the magnitude bins and the total number of bins, $N_\textrm{features}$. Recall from Eq.~\eqref{eq:n_M} that $N(M)$ is the cumulative event count of our {\em entire} training catalog, and that $N_\textrm{tot} \approx 3.8\times 10^5$. In this work, we select $\alpha=0.7$, and this leads to $N_\textrm{feature} = 35$ non-empty bins. We emphasize that the characteristic magnitudes $M_j$ are derived from the entire training catalog, independent of window index $i$.

We now define the feature vector $\mathbf{X}_i$ for each window $i$ from the sub-catalog of event magnitudes $m^{(i)}_k$, where $k$ indexes the events within the window. We test two simple schemes for $\mathbf X_i$, sensitive to the counts and amplitudes in the catalog. Mathematically, the $j$th components of these two feature vectors are
\begin{align}
X^{\textrm{count}}_{i,j} &=\sum_{m^{(i)}_k \geq M_j} 1 \\
X^{\textrm{ampl}}_{i,j} &= \sum_{m^{(i)}_k \geq M_j} 10^{m^{(i)}_k}  \label{eq:ampl_features}
\end{align}
where the sums run over all sub-catalog events $k$ with magnitude $m^{(i)}_k$ at least $M_j$. In words,  $X_{i,j}^\textrm{count}$ is the count of events in the $i$th window with magnitude exceeding $M_j$ (analogous to $N(M_j)$, but restricted to sub-catalog $i$). The features $X_{i,j}^\textrm{ampl}$ measure the total amplitude of all events in window $i$ above the magnitude $M_j$.

\subsection{Catalog Ablation Test}

To understand how our catalog-RF model performs with decreasing information, we vary the magnitude of completeness of our catalog by discarding all events with magnitude below a cutoff,  $M_\textrm{cut}$. In practice, we implement this ablation by removing features  $X_{i,j}$ for indices  $j$ that correspond to magnitudes $M_j$ below the cutoff $M_{\textrm{cut}}$. We can then quantify the performance as a function of the ablation cutoff $M_\textrm{cut}$.

\section{Results}
\subsection{Performance on Full Catalogs}

\begin{figure}[h]
 \centering
 \includegraphics[width=32pc]{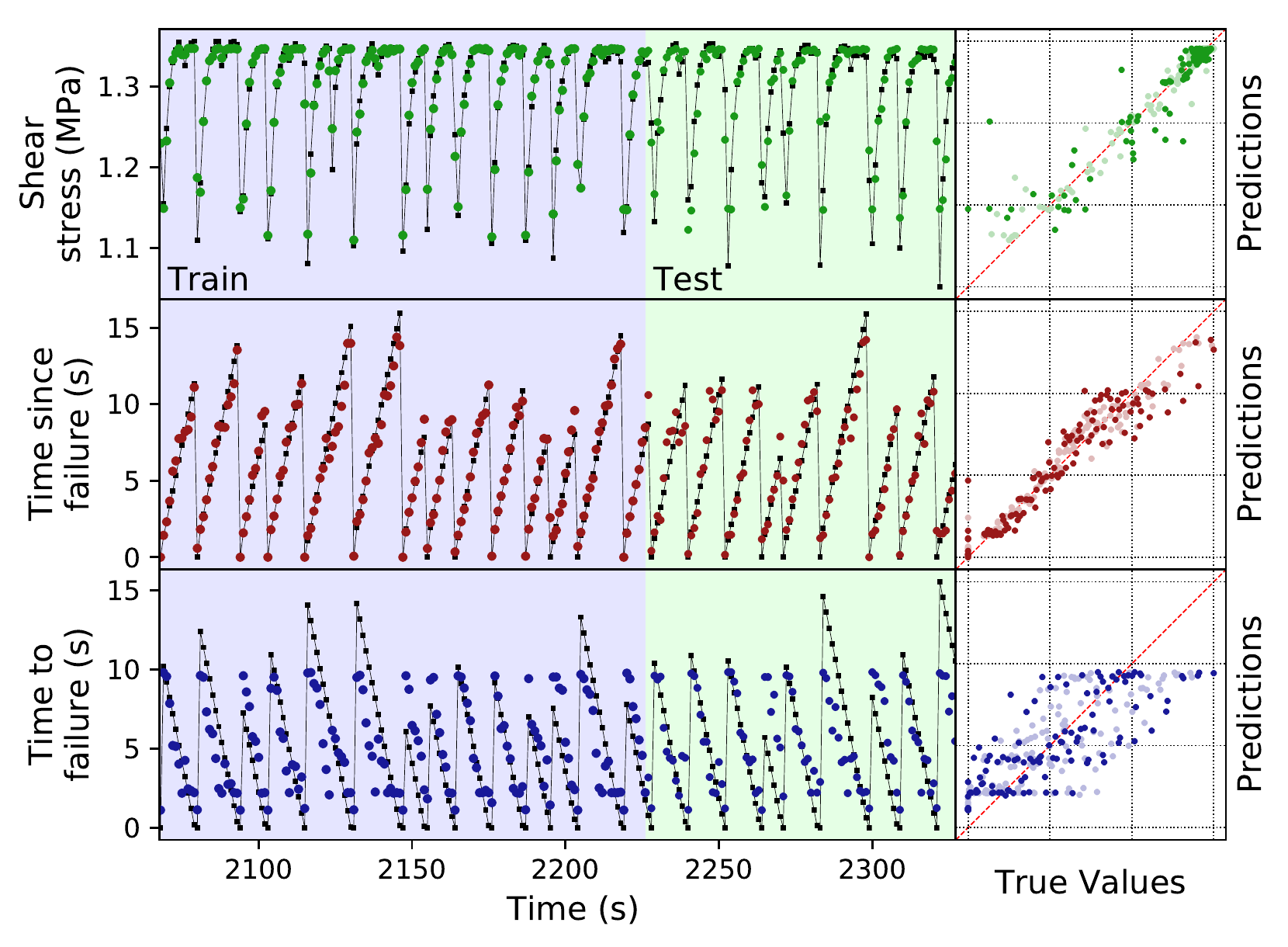}

 \caption{Regression performance for the catalog-ML model using amplitude-based features. The rows show performance for shear stress, TSF, and TTF, respectively. The left panel of each row shows predictions on training and testing data over time. True values are shown as black squares, and predicted values as colored circles. Time windows for features are contiguous and non-overlapping. Each point is plotted at the end of the time window represented.  The right panel of each row plots the prediction vs. true values for each label. Testing data is shown with dark markers, and training data is shown with faint markers.}
 \label{figregression}
 \end{figure}

Table \ref{tabregression} shows the $R^2$ performance of our catalog-RF models for shear stress, TSF, and TTF, trained using the full catalogs.  The TTF models ($R^2 = 0.551, R^2=0.612$) do not capture fine details. The TTF prediction is worse than the TSF and shear stress models ($R^2 > 0.8$)---in other words, it is harder to predict the future than to estimate aspects of the past or present. Table \ref{tabregression} also indicates that the cumulative counts  $\mathbf{X}^\textrm{count}$ are better for predicting the shear stress, and the cumulative amplitudes  $\mathbf{X}^\textrm{ampl}$ are slightly better for TTF and TSF. The count-based catalog features predict shear stress with an accuracy of $R^2 = 0.898$, nearly equal to the continuous acoustic approach of \citet{BRLFriction}, who report $R^2 = 0.922$ using the same data (that is, same train and test segments on the same experiment). We note that although the same data are analyzed, the methods cannot be compared exactly; for example, the time window in \citet{BRLFriction} is $1.33$ seconds, and each time window has $90\%$ overlap with the previous window. (Recall that in our work the time window length is $1$ second, and windows do not overlap.)  Figure~\ref{figregression} shows the predictions compared to the true labels over time for the amplitude-based features $\mathbf{X}^\textrm{ampl}$, as well as scatter-plots showing a comparison of the true and predicted labels. Figure~S1 shows that models perform nearly as well for shear stress and TSF when analyzing a larger region of 573 s of data (44 failure cycles), but with reduced performance for TTF. This fact can be accounted for by drift in cycle length over time; TTF performance is restored by randomly assigning cycles to testing data from throughout the entire data analysis region (Fig.~S2).

\begin{table}[h]
 \centering
 \begin{tabular}{l  r  r  r }
 \toprule
   & Shear Stress & TSF  & TTF \\
 \midrule
  $\mathbf{X}^\textrm{count}$ & 0.898 & 0.842 & 0.551 \\
  $\mathbf{X}^\textrm{ampl}$ & 0.848 & 0.882 & 0.612 \\
  \bottomrule
 \end{tabular}
 \caption{$R^2$  performance for regression of shear stress, TSF, and TTF for catalog-RF models on complete catalogs. The choice of count or amplitude based features ( $\mathbf{X}^\textrm{count}$ or  $\mathbf{X}^\textrm{ampl}$) affects performance slightly.}
 \label{tabregression}
\end{table}

\subsection{Performance on Ablated Catalogs}

 \begin{figure}[h]
 \centering
 \includegraphics[width=32pc]{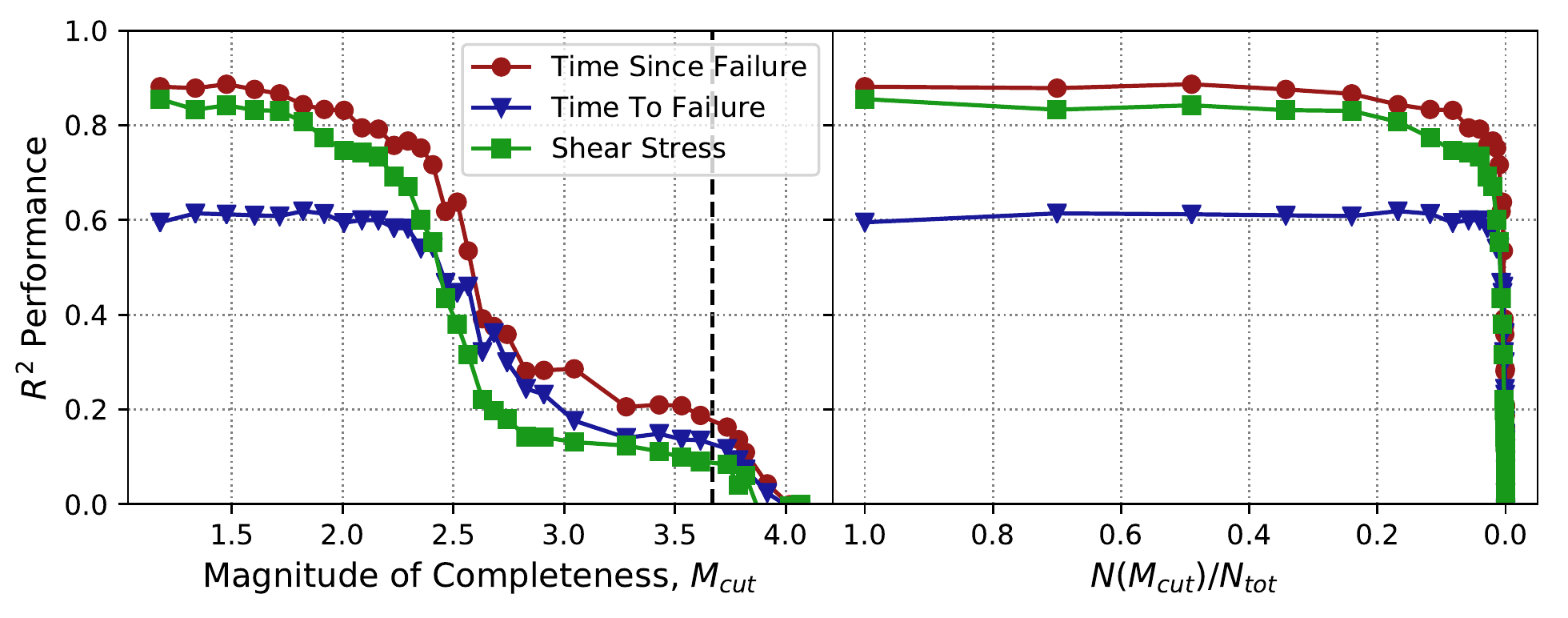}

 \caption{The accuracy of the catalog-RF model decreases as training sub-catalogs are artificially limited to varying magnitudes of completeness, $M_\textrm{cut}$. Left: $R^2$ performance vs. $M_{\textrm{cut}}$. The drop of $R^2$ for $M_\textrm{cut}$ between 2.0 and 3.0 indicates that this range of event magnitudes plays a key role in our RF models. The dashed vertical line at $M_{\textrm{cut}} = 3.67$ denotes the smallest event magnitude associated with a large stress drop. Right: $R^2$ is plotted against $N(M_\textrm{cut})/N_\textrm{tot}$, i.e. the fraction of events remaining in the full catalog after ablation. One observes that most events contribute very little to the catalog-RF performance.}
 \label{figablation}
 \end{figure}

Next we build our catalog-RF models on ablated catalogs that include events only above an imposed magnitude of completeness, $M_\textrm{cut}$. For this study, we use the amplitude features $\mathbf{X}^\textrm{ampl}$ of Eq.~\eqref{eq:ampl_features}.  Figure~\ref{figablation}, left side, shows the $R^2$ performance as a function of the imposed magnitude cutoff.  As one might expect, catalog-RF models generally perform worse when provided less information. However, it is interesting to note that $R^2$ decays relatively slowly until $M_\textrm{cut}$ reaches about 2.0, after which the performance drops precipitously, plateauing near magnitude 3.0. In this regard we say that $M_{\textrm{cut}} = 2.0$ provides a {\em sufficient magnitude of completeness} for the catalog-RF model with the given window size.  In Fig.~\ref{figablation}, right side, we plot $R^2$ as a function of the remaining event fraction $N(M_\textrm{cut})/N_\textrm{tot}$. This highlights that the performance drop at $M_{\textrm{cut}} \approx 2.0$ corresponds to ablating $\approx 80\%$ of the events from the catalog. Likewise, the sharp decay in performance indicates that events $M \lesssim 2.5$ are {\em required} to make an accurate determination of the fault state. The performance of the method for $M_{\textrm{cut}} \gtrsim 3 $ plateaus at a non-zero value of $R^2$ because the algorithm is still able to differentiate between windows that contain a large failure and those that do not, but is otherwise unable to discern useful differences between sub-catalogs. 

A study of the ablation curve as a function of window size (Fig.~S3) shows that the sharp decay in performance shifts to smaller cutoff magnitudes as the window size is decreased; while regression can be performed with small window sizes, the sufficient magnitude of completeness is lower. This leads us to hypothesize that the sharp decay in performance may be related to the number of events within a window; to keep the number of events in a window fixed, one must lower the magnitude of completeness along with the window size. 

\section{Conclusions}

We analyze a dense acoustic event catalog obtained from biaxial shearing friction experiment.  The device produces dozens of aperiodic stick-slip events during an experiment. Stacking catalog statistics over many cycles shows that as failure approaches, events are more common and typically larger (Fig.~\ref{figstack}).  We then use a machine learning workflow to show that physical characteristics of the fault (shear stress, TSF, and TTF) can be obtained from catalog statistics over a short time window using a Random Forest regression model. Previous work employed the continuous AE, and showed remarkable ability to forecast failure \citep{rouet17ML}, as well as to infer the instantaneous shear stress and friction \citep{BRLFriction}.  Here, we demonstrate that the continuous acoustic waveform is not needed if a catalog {\em with sufficient completeness} is available. Our models achieve similar accuracy to the continuous approach.

We note, however, that our catalogs are extraordinarily well resolved. To create these catalogs, we recorded AE at very high frequency (4 MHz) using a sensor very near the fault (distance $\approx 22$ mm). One would not expect such fidelity in real earthquake catalogs, except perhaps if the sensors happened to be located very close to the slip patch.  We find that the smallest events in our catalog contribute little to the prediction accuracy; increasing $M_\textrm{cut}$ from 1.0 to 2.0 only slightly diminishes the catalog-RF model performance (visible in Fig.~\ref{figablation}).

The fact that the smallest events are not {\em required} offers hope that a catalog-based ML approach could eventually contribute to the understanding of natural seismicity. We note a few potential differences between laboratory conditions and natural seismicity that may pose challenges. Laboratory data corresponds to a single isolated fault. Tectonic fault zones are typically comprised of many sub-faults, and predictions may be vastly more difficult. While precursors are frequently observed in laboratory studies, they are not as reliably observed in natural faults. Moreover, large seismic cycles in the earth are far slower \citep{nelson2006,Ellsworth2013}, and thus fewer cycles are available to train ML models. 

We note that \citet{RouetCascadia18} have applied the continuous approach developed in the laboratory to slow slip in Cascadia with surprisingly good results.  Thus we are hopeful that the catalog-based approach described here may prove fruitful to the study of real earthquakes, given sufficient catalog quality. It may have particular importance in regions where continuous data are not yet available, where the continuous data is quite noisy, or for examining archival catalogs.

\acknowledgments
Work supported by institutional support (LDRD) at the Los Alamos National Laboratory.  We gratefully acknowledge the support of the Center for Nonlinear Studies. CM was supported by NSF-EAR1520760 and DE-EE0006762. We thank Andrew Delorey, Robert A. Guyer, Bertrand Rouet-Leduc, Claudia Hulbert, and James Theiler for productive discussions. The data used here are freely available from the Penn State Rock Mechanics lab at \url{http://www3.geosc.psu.edu/~cjm38/}.

\bibliography{references.bib}

\end{document}


\supportinginfo{Identification of lab-quakes and fault state from catalogs with limited magnitudes of completeness}

\section*{Contents}

\begin{enumerate}
\item Text S1: Experimental details
\item Text S2: Catalog generation from acoustic emissions
\item Text S3: Random forest details
\item Table S1: Performance of various regression algorithms
\item Figure S1: Regression for long time segment with contiguous train/test split
\item Figure S2: Regression for long time segment with random cycle train/test split
\item Figure S3: Ablation analysis for varying time-window sizes
\end{enumerate}

\section*{Text S1: Experimental details}

We constructed samples for the double direct shear (DDS) assembly following the procedures detailed in previous works \citep{Anthony05,riviere2018}. Steel guide plates are mounted on the side blocks of the DDS to prevent gouge loss on the front and back edges. The faces of the forcing blocks in contact with the gouge layers are grooved to ensure that shear occurs within the layer; grooves are 0.8 mm deep and spaced every 1mm perpendicular to the direction of shear. In addition, a small rubber jacket covers the bottom half of the sample to minimize material loss throughout the experiment. Forces on faults are measured with strain-gauge load cells placed in series with the horizontal and vertical pistons. Displacements on the fault are measured using direct current displacement transducers that are coupled to the horizontal and vertical pistons. Stress and displacement data are measured continuously at 1kHz throughout the experiment from a 24-bit recording system. All measurements of shear stress and displacement for the DDS configuration are relative to one fault. In addition, we have instrumented 36 broad-band piezoceramic sensors (~0.02-2 MHz) into steel 10 x 10 cm\textsuperscript{2} blocks. The sensors are in blind holes 2-mm from the block face and the block is placed adjacent to the fault zone to monitor acoustic emission activity \citep{riviere2018}. The acoustic emission data analyzed here are recorded continuously at 4 MHz from a single piezoceramic sensor using a 14-bit Verasonics data acquisition system.

\section*{Text S2: Catalog generation from acoustic emissions}

We use a simple event detection algorithm to construct a catalog from the acoustic emission time series signal.  The algorithm has been designed to account for both the response of the sensor and the background noise level. The first step is to compute the envelope of the signal and use a moving average to smooth it over the time scale $\tau_\textrm{smooth}$. From this smoothed envelope we examine peaks and label them as events if they satisfy the following criteria: 1) The peak is above a minimum amplitude threshold, $A_\textrm{min}$, related to the average background noise. 2) The peak is separated from previous peaks by a minimum inter-event time threshold, $\tau_\textrm{min}$. 3) The peak does not occur during in the coda of an event -- this is defined using an exponential ring-down function computed over the ten previous events. Mathematically, event $k$ is not part of a coda if the peak amplitude $A(t_k)$ satisfies
 \begin{equation}
A(t_k) > A(t_{k-i})\exp{\frac{-{(t_k-t_{k-i})}}{\tau_\textrm{ring-down}}}
\label{eq:A_t0}
\end{equation}
for $i = 1...10$.  The parameters of this algorithm are selected by evaluating the catalog quality on random 5~$\mu s$ snapshots of the acoustic signal using trial parameters and visual evaluation of catalog quality. A high-quality catalog maximizes the number of distinct visual phenomena selected as events, while at the same time minimizes the assignment of multiple events to a single phenomenon. In the end, we selected $A_\textrm{min} = 15$, $\tau_\textrm{smooth} = 7.75$ {\textmu}s,  $\tau_\textrm{min} = 150$ {\textmu}s, and $\tau_\textrm{ring-down} = 156.7$ {\textmu}s.

\section*{Text S3: Random forest details}
We employed 100 decision trees per random forest, each of which sees a bootstrap resampling of the training set, with a mean-squared-error cost function. Before training a random forest model (i.e., designing each tree's yes/no questions from the data), one may specify the maximum depth of the decision trees, which controls the allowable complexity of the random forest model. (If the maximum depth is larger, the random forest can fit more complicated datasets, but also becomes more prone to overfitting irrelevant details of the data.) For each regression task, we used 10-fold cross-validation of the training set to select the maximum depth; the possible depths are 1, 2, 4, 6, 8, and 12. The folds for cross-validation are selected by randomly partitioning the training set. Cross-validation uses the $R^2$ score as the criterion for the best model.  After identifying the tree depth which cross-validates most successfully, we retrain a forest with that maximal depth to the entire training set. Other regularization hyperparameters such as the minimal leaf size and random feature sub-selection were not utilized.

\begin{table}[h]
 \centering
\begin{tabular}{llrrrrr}
\toprule
           &       &    RFR &    GBR &  ElasticNet &  Lasso &  Ridge \\
Regression task & Feature vector &        &        &               &          &          \\
\midrule
Shear stress & $\mathbf X^\textrm{count}$ &  {\bf 0.901} &  0.893 &         0.715 &    0.715 &    0.800 \\
           & $\mathbf X^\textrm{ampl}$ &  {\bf 0.860} &  0.827 &         0.817 &    0.817 &    0.796 \\
TSF & $\mathbf X^\textrm{count}$ &  0.843 &  \bf{0.886} &         0.630 &    0.634 &    0.743 \\
           & $\mathbf X^\textrm{ampl}$ &  0.888 &  \bf{0.891} &         0.779 &    0.779 &    0.771 \\
TTF & $\mathbf X^\textrm{count}$ &  0.547 &  \bf{0.579} &         0.533 &    0.533 &    0.571 \\
           & $\mathbf X^\textrm{ampl}$ &  \bf{0.618} &  0.578 &         0.588 &    0.588 &    0.471 \\
\bottomrule
\end{tabular}
\caption{$R^2$ performance for various regression models. We compare random forest regression (RFR) \citep{Breiman2001}, gradient boosting regression (GBR) \citep{Friedman2001}, and linear regression with ElasticNet \citep{hui2005}, lasso \citep{tibshirani96}, and ridge \citep{hoerl1970} regularization penalties as implemented in scikit-learn \citep{pedregosa2011scikit}. The best performing model for each regression task and feature vector type is shown in bold. RFR and GBR produce models of similar quality and both outperform linear regression schemes.}
 \label{methodchoices}
\end{table}

\begin{figure}[h]
 \centering
 \includegraphics[width=32pc]{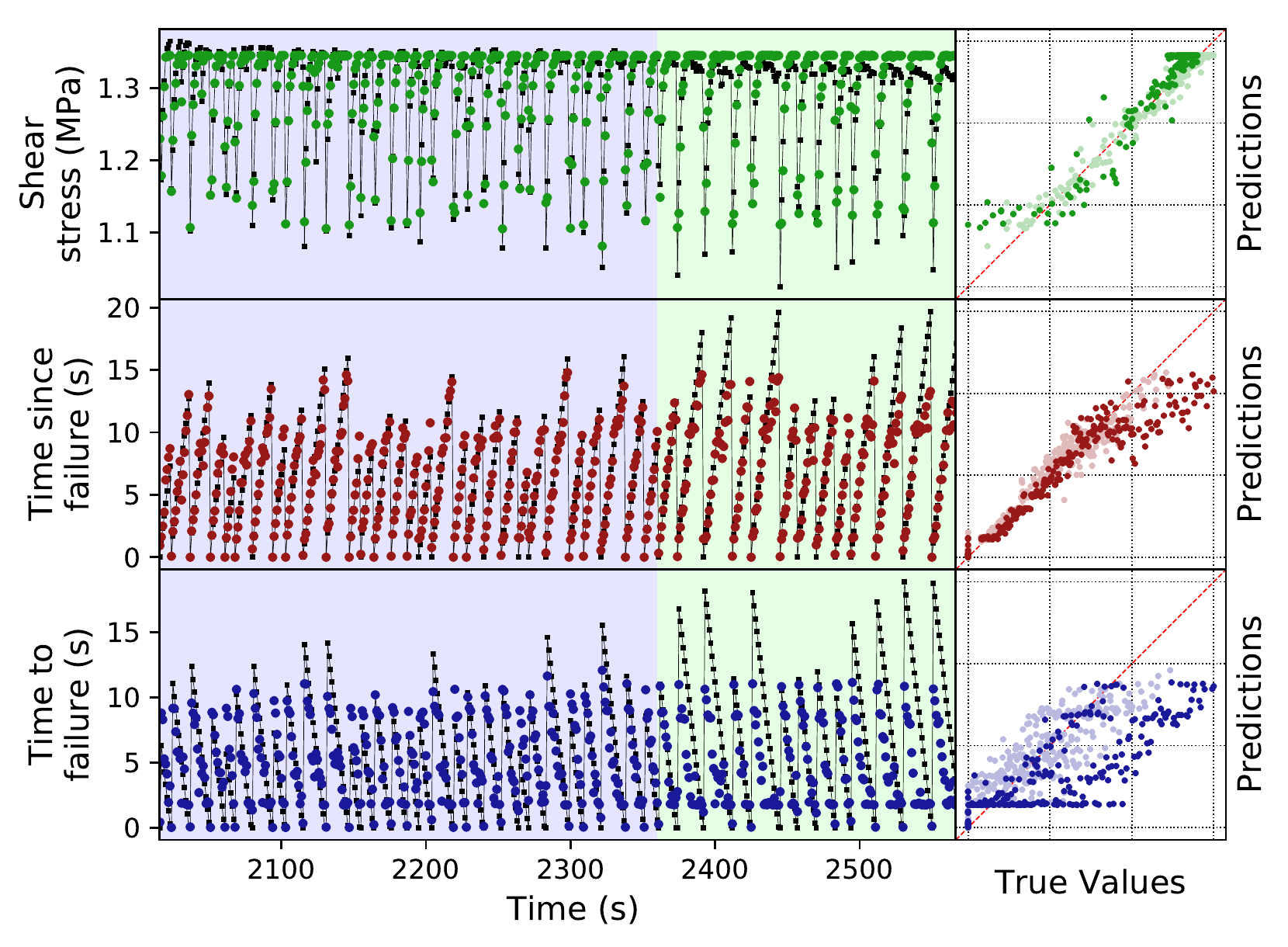}

 \caption{Regression performance for the catalog-ML model using amplitude-based features using 573 seconds of data. The rows show performance for shear stress ($R^2 =0.869$) , TSF ($R^2 =0.813$), and TTF ($R^2 =0.284$), respectively. The left panel of each row shows predictions on training (region shaded blue) and testing (region shaded green) data over time. True values are shown as black squares, and predicted values as colored circles. The right panel of each row plots the prediction vs. true values for each label. Testing data is shown with bold markers, and training data is shown with faint markers.}
 \label{figregression}
 \end{figure}

\begin{figure}[h]
 \centering
 \includegraphics[width=32pc]{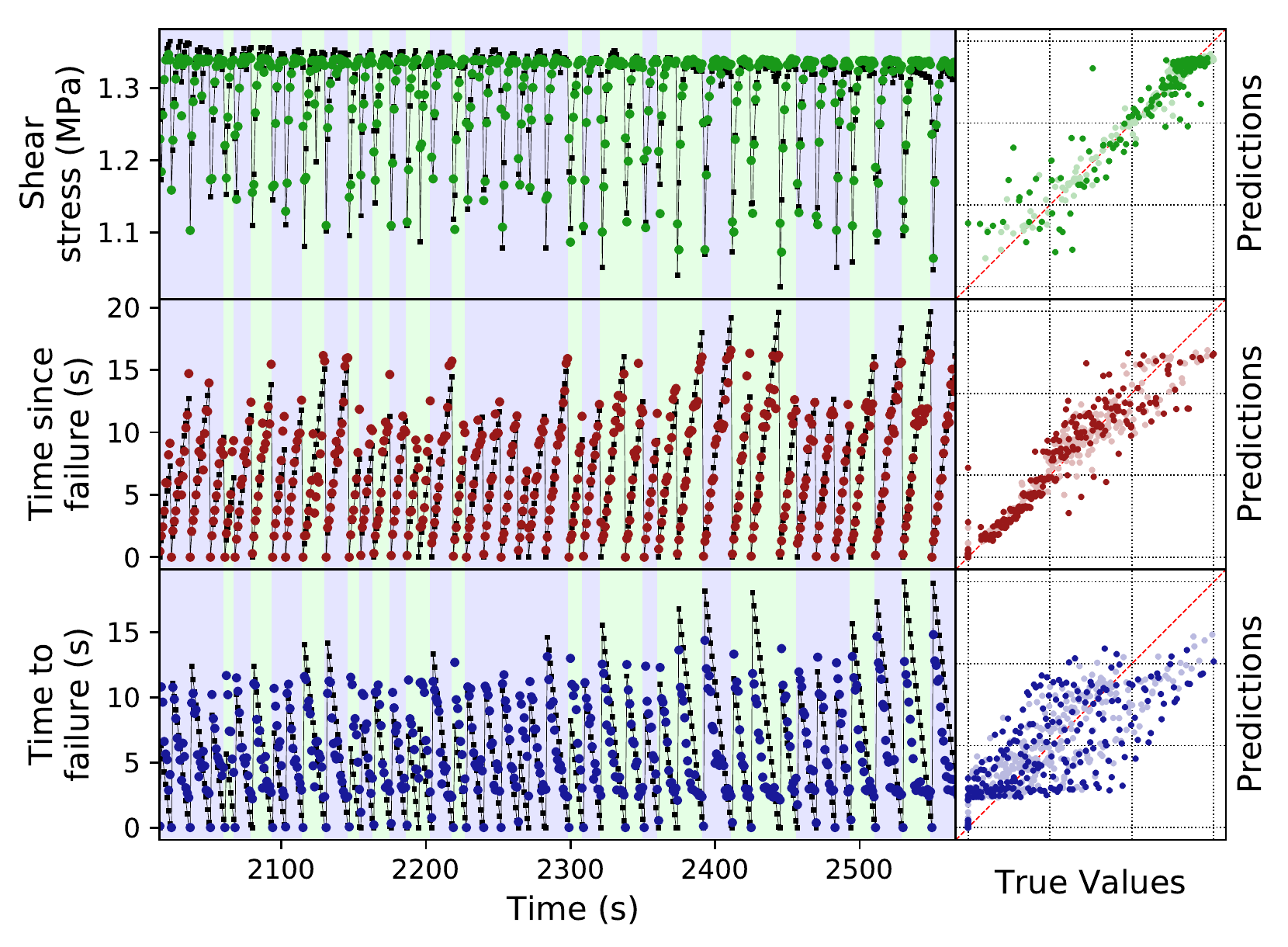}

 \caption{Regression performance for the catalog-ML model using amplitude-based features using 573 seconds of data, where each failure cycle is individually assigned to either training or testing data. The rows show performance for shear stress ($R^2 =0.881$), TSF ($R^2 =0.869$), and TTF ($R^2 =0.511$), respectively. The left panel of each row shows predictions on training (region shaded blue) and testing (region shaded green) data over time. True values are shown as black squares, and predicted values as colored circles. The right panel of each row plots the prediction vs. true values for each label. Testing data is shown with bold markers, and training data is shown with faint markers.}
 \label{figregression}
 \end{figure}
\begin{figure}[h]
 \centering
 \includegraphics[width=20pc]{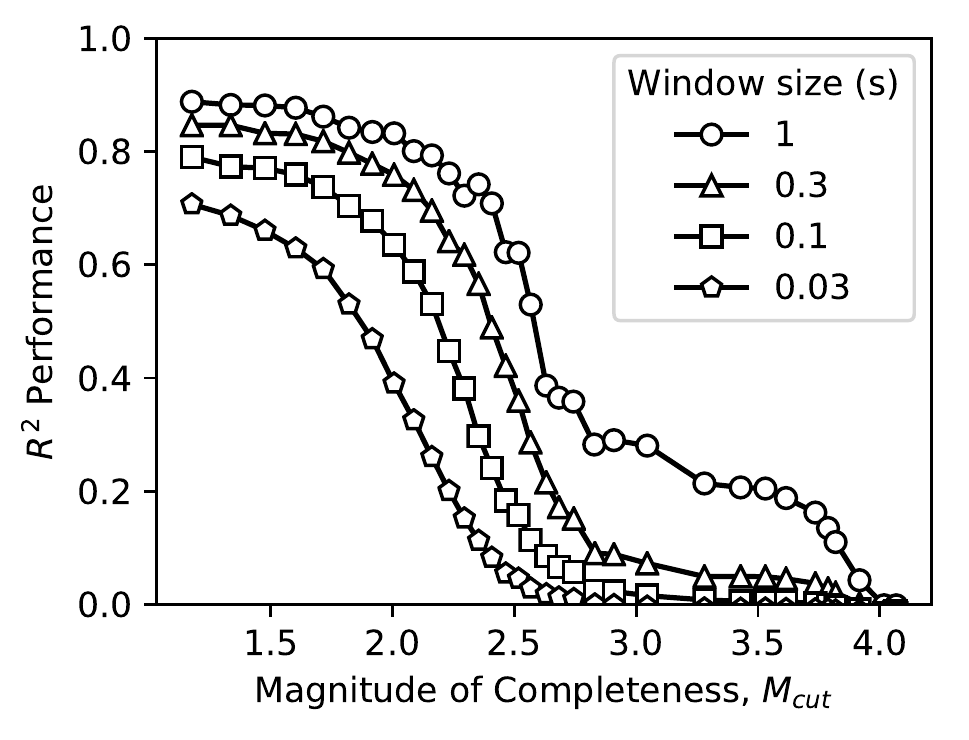}

 \caption{Ablation analysis for time since failure (TSF) showing $R^2$ performance vs. artificial magnitude of completeness for different window sizes. As the window size decreases, performance decreases. Furthermore, the curve shifts left; the magnitude of completeness necessary to achieve a given $R^2$ performance lowers along with window size. This figure demonstrates that while regression can be performed with small window sizes, the process relies on having information about progressively smaller events.}
 \label{fig:windowsizeabalation}
 \end{figure}

\bibliography{references.bib}